\documentclass{elsart}

\usepackage{graphicx}

\usepackage{amssymb}

\usepackage{mathptmx}
\usepackage{url}
\newcommand{\q}[2]{\ensuremath{#1\ \mathrm{#2}}}
\newcommand{\psip}{\ensuremath{\psi (2S)}}
\newcommand{\pbarp}{\ensuremath{\bar{p} p}}
\newcommand{\ee}{\ensuremath{e^+ e^-}}
\newcommand{\jpsi}{\ensuremath{J/\psi}}
\newcommand{\psinc}{\ensuremath{\jpsi + X}}

\newcommand{\Gin}{\ensuremath{\Gamma_{\mathrm{in}}}}
\newcommand{\Gout}{\ensuremath{\Gamma_{\mathrm{out}}}}
\newcommand{\Gee}{\ensuremath{\Gamma_{\ee}}}
\newcommand{\Gpp}{\ensuremath{\Gamma_{\pbarp}}}
\newcommand{\GGG}{\ensuremath{\Gee \Gpp / \Gamma}}
\newcommand{\GiGoG}{\ensuremath{\Gin \Gout / \Gamma}}
\newcommand{\Lum}{\ensuremath{\mathcal{L}}}
\newcommand{\lik}{\ensuremath{\Lambda}}
\newcommand{\sBW}{\ensuremath{\sigma_\mathrm{BW}}}
\newcommand{\sBWr}{\ensuremath{\sigma_\mathrm{BWr}}}
\newcommand{\scf}{\ensuremath{b}}
\newcommand{\sbkg}{\ensuremath{\sigma_\mathrm{bkg}}}

\newcommand{\fcav}{\ensuremath{f^\mathrm{cav}}}
\newcommand{\frf}{\ensuremath{f^\mathrm{rf}}}
\newcommand{\vrf}{\ensuremath{v^\mathrm{rf}}}
\newcommand{\betarf}{\ensuremath{\beta^\mathrm{rf}}}
\newcommand{\gammarf}{\ensuremath{\gamma^\mathrm{rf}}}
\newcommand{\wrf}{\ensuremath{w^\mathrm{rf}}}
\newcommand{\frfz}{\ensuremath{f^\mathrm{rf}_0}}
\newcommand{\frfi}{\ensuremath{f^\mathrm{rf}_i}}
\newcommand{\dL}{\ensuremath{\Delta L}}
\newcommand{\reff}{\ensuremath{\varepsilon^X_\mathrm{co}/\varepsilon^X_\mathrm{cf}}}
\newcommand{\muee}{\ensuremath{\mu^{ee}}}
\newcommand{\Nee}{\ensuremath{N^{ee}}}
\newcommand{\effee}{\ensuremath{\varepsilon^{ee}}}
\newcommand{\muX}{\ensuremath{\mu^{X}}}
\newcommand{\NX}{\ensuremath{N^{X}}}
\newcommand{\effXcf}{\ensuremath{\varepsilon^{X}_\mathrm{cf}}}
\newcommand{\resG}{\q{290 \pm 25 \mathrm{(sta)} \pm 4 \mathrm{(sys)}}{keV}}
\newcommand{\resA}{\q{579 \pm 38 \mathrm{(sta)} \pm 36 \mathrm{(sys)}}{meV}}

\newcommand{\code}[1]{\textsc{#1}}

\begin{document}

\begin{frontmatter}



\title{Precision measurements of the total and partial widths
of the \protect\psip\ charmonium meson
with a new complementary-scan technique in \protect\pbarp\ annihilations}



\begin{center}
Fermilab E835 Collaboration
\end{center}

\author[fe]{M.~Andreotti,}
\author[ge,to]{S.~Bagnasco,}
\author[fe]{W.~Baldini,}
\author[fe]{D.~Bettoni,}
\author[to]{G.~Borreani,}
\author[ge]{A.~Buzzo,}
\author[fe]{R.~Calabrese,}
\author[to]{R.~Cester,}
\author[fe]{G.~Cibinetto,}
\author[fe]{P.~Dalpiaz,}
\author[fnal]{G.~Garzoglio,}
\author[fnal]{K.~E.~Gollwitzer,}
\author[mn]{M.~Graham,}
\author[fnal]{M.~Hu,}
\author[nw]{D.~Joffe,}
\author[nw]{J.~Kasper,}
\author[ir]{G.~Lasio,}
\author[ge]{M.~Lo~Vetere,}
\author[fe]{E.~Luppi,}
\author[ge]{M.~Macr\`{\i},}
\author[ir]{M.~Mandelkern,}
\author[to]{F.~Marchetto,}
\author[ge]{M.~Marinelli,}
\author[to]{E.~Menichetti,}
\author[nw]{Z.~Metreveli,}
\author[fe,to]{R.~Mussa,}
\author[fe]{M.~Negrini,}
\author[mn,to]{M.~M.~Obertino,}
\author[ge]{M.~Pallavicini,}
\author[to]{N.~Pastrone,}
\author[ge]{C.~Patrignani,}
\author[fnal]{S.~Pordes,}
\author[ge]{E.~Robutti,}
\author[ir,nw]{W.~Roethel,}
\author[nw]{J.~Rosen,}
\author[nw]{P.~Rumerio,}
\author[mn]{R.~W.~Rusack,}
\author[ge]{A.~Santroni,}
\author[ir]{J.~Schultz,}
\author[mn]{S.~H.~Seo,}
\author[nw]{K.~K.~Seth,}
\author[fnal,fe]{G.~Stancari,\corauthref{gs}}
\author[fe,ir]{M.~Stancari,}
\author[nw]{A.~Tomaradze,}
\author[nw]{I.~Uman,}
\author[mn]{T.~Vidnovic,}
\author[fnal]{S.~Werkema,}
\author[nw]{P.~Zweber}

\address[fnal]{Fermi National Accelerator Laboratory, Batavia, IL 60510, USA}
\address[fe]{Istituto Nazionale di Fisica Nucleare and University of Ferrara,
44100 Ferrara, Italy}
\address[ge]{Istituto Nazionale di Fisica Nucleare and University of Genova,
16146 Genova, Italy}
\address[ir]{University of California, Irvine, CA 92697, USA}
\address[mn]{University of Minnesota, Minneapolis, MN 55455, USA}
\address[nw]{Northwestern University, Evanston, IL 60208, USA}
\address[to]{Istituto Nazionale di Fisica Nucleare and University of Torino,
10125 Torino, Italy}

\corauth[gs]{Corresponding author:
             Dr.~Giulio Stancari,
             Istituto Nazionale di Fisica Nucleare,
             Via Saragat 1,
             I-44100 Ferrara FE,
             Italy.
             Phone: +39.0532.974330;
             fax: +39.0532.974343;
             e-mail: \url{stancari@fe.infn.it}.}

\begin{abstract}
We present new precision measurements of the \psip\ total and partial widths from excitation curves obtained in antiproton-proton annihilations
by Fermilab experiment E835 at the Antiproton Accumulator in the year~2000.
A new technique of complementary scans
was developed to study narrow resonances with stochastically cooled antiproton beams.
The technique relies on precise revolution-frequency and orbit-length measurements,
while making the analysis of the excitation curve almost independent of machine lattice parameters.
We study the \psip\ meson through the processes
\( \pbarp \to \ee \) and
\( \pbarp \to \jpsi + X \to \ee + X \).
We measure the width to be
\( \Gamma = \resG \)
and the combination of partial widths
\( \Gee \Gpp / \Gamma = \resA \),
which represent the most precise measurements to date.
\end{abstract}


\end{frontmatter}

%
%
\section{Introduction}

A precise measurement of the excitation curve of narrow charmonium
resonances depends both on the detection technique
(event statistics, detector efficiency)
and on the properties of the beam-energy spectrum.  In \ee\
annihilations, the
beam-energy spread is substantially larger than the resonance width.  The
BES Collaboration at BEPC published two measurements of the
\psip\ width in \ee\ collisions at center-of-mass energies between 3.67~GeV and 3.71~GeV~\cite{bai:2002,ablikim:2006}.  The combination \GGG\ was
recently measured by the BABAR Collaboration at PEP-II using initial-state radiation~\cite{aubert:2006}.  In
\pbarp\ annihilations, the event statistics are lower, but one can
take advantage of stochastically cooled antiproton beams, with FWHM
energy spreads of 0.4--0.5~MeV in the center-of-mass frame,
to measure the width directly from the excitation curve generated by scanning the beam across the resonance.
Fermilab experiment E760 measured the widths of the \jpsi\ and \psip\
mesons~\cite{armstrong:1993}.  The uncertainty was dominated by event
statistics and statistical fluctuations in the beam position
measurements.  A sizeable systematic uncertainty was due to the measurement of the beam-energy spectrum.  In this paper, we present
results obtained in \pbarp\ annihilations by Fermilab experiment E835
from data collected during the year 2000 run.  A new scanning
technique, together with higher event statistics, improvements in the
beam position measurement and momentum-spread analysis, allow us to reach the
highest precision to date.

\section{Experimental technique}

The main features of the experiment are summarized here.
A full description can be found in Ref.~\cite{garzoglio:2004}.

In experiment E835, antiprotons circulating in the Antiproton
Accumulator intersect an internal hydrogen gas-jet target.  The beam
is cooled and decelerated to scan charmonium resonances.  The
operation of the Accumulator for E835 is described in
Ref.~\cite{mcginnis:2003}.

The E835 detector is a nonmagnetic spectrometer designed to extract,
from a large hadronic background, electron-positron pairs of high
invariant mass as a signature of charmonium formation.
The apparatus has full acceptance in azimuth, with a cylindrical central system and a planar forward system.
The central detector system includes three segmented hodoscopes,
straw-tube and scintillating-fiber trackers,
a threshold gas Cherenkov counter,
and an electromagnetic lead-glass calorimeter.
In the forward direction, a veto counter and a planar lead-glass electromagnetic calorimeter are placed.

The event selection is described in detail in our paper
on \psip\ branching ratios~\cite{andreotti:2005}.
The main hardware trigger requires two charged tracks, each defined by a coincidence between two hodoscope counters, with at least one of the two particles tagged as an electron or positron by a signal in the corresponding Cherenkov cell.
In addition, two energy deposits that are roughly back-to-back in azimuth are required in the central calorimeter, with an invariant mass greater than a given fraction of the center-of-mass energy.
A preliminary off-line selection requires that all \ee\ candidates have an invariant mass greater than 2.6~GeV.
A maximum-likelihood method called `electron weight' rejects backgrounds, mainly photon conversions and Dalitz decays of the pion, that mimic electron or positron tracks in the detector.
It is based on pulse height in the hodoscopes and in the Cherenkov counter, and on the shape of the electromagnetic shower in the central calorimeter.
The processes
\( \pbarp \to \ee \) and
\( \pbarp \to \jpsi + X \to \ee + X \)
are finally selected using kinematic fits requiring a $\chi^2$ probability greater than~$10^{-4}$.
The overall efficiency,
including detector acceptance, hardware trigger, and off-line selections,
is about 40\%
(see Section~\ref{sec:results}), while background contamination is only 0.1\% for the \ee\ channel and 1\% for the inclusive channel.

Two scans of the \psip\ resonance were performed,
in January~2000 (47~hours of data taking) and in June~2000 (21~hours).
For each run~$i$, the luminosity~$\Lum_i$
and the number of selected events~$N_i$
are shown in Tables~\ref{tab:stack01} and~\ref{tab:stack29}.

The resonance parameters are determined from a maximum-likelihood fit
to the excitation curve.
For each data-taking run (subscript~$i$),
we assume that the average number of observed events~$\mu_i$
in each channel depends on a Breit-Wigner cross section \sBWr\
and on the center-of-mass energy distribution, $B_i$, as follows:
\begin{equation}
  \mu_i = \Lum_i \left[ \varepsilon_i \int \sBWr(w) \, B_i(w) \, dw
    + \sbkg
  \right],
  \label{eq:mu}
\end{equation}
where~$w$ is the center-of-mass energy,
$\varepsilon_i$ is the detector efficiency,
$\Lum_i$ is the integrated luminosity,
and \sbkg\ is a constant background cross section.
The integral is extended over the energy acceptance of
the machine.
The spin-averaged Breit-Wigner cross section
for a spin-$J$ resonance of mass~$M$ and width~$\Gamma$
formed in \pbarp\ annihilations is
\begin{equation}
  \sBW(w) = \frac{(2J+1)}{(2S+1)^2}
  \frac{16\pi}{w^2-4m^2}
  \frac{(\Gin \Gout / \Gamma) \cdot \Gamma}
  {\Gamma^2 + 4 (w-M)^2};
\end{equation}
$m$ and $S$ are the (anti)proton mass and spin,
while~$\Gin$ and~$\Gout$ are the partial resonance widths for the entrance (\pbarp, in our case) and exit channels.
The Breit-Wigner cross section
is corrected for initial-state radiation to obtain
\sBWr~\cite{armstrong:1993,kennedy:1992}:
\begin{eqnarray}
  \sBWr(w) & = & \scf \int_0^{w/2}
  \frac{dk}{k}
  \left(\frac{2k}{w}\right)^\scf
  \sBW(\sqrt{w^2-2kw}) \\
  & = & \left(2/w\right)^\scf
  \int_0^{(w/2)^\scf} dt \, \sBW(\sqrt{w^2-2t^{1/\scf}w}),
\end{eqnarray}
where the second form is more suitable for numerical integration and
$\scf(w)$ is the semiclassical collinearity factor~\cite{kennedy:1992},
equal to 0.00753 at the \psip.

The resonance mass~$M$, width~$\Gamma$, `area'~$(\GiGoG)$ and the
background cross section~\sbkg\ are left as free parameters in the
maximization of the log-likelihood function
\( \log(\lik) = \sum_i \log P(\mu_i,N_i) \),
where~$P(\mu,N)$ are Poisson
probabilities of observing~$N$ events when the mean is~$\mu$.

Both channels \( \pbarp \to \ee \) and \( \pbarp \to \jpsi
+ X \to \ee + X \) are fit simultaneously to the same mass and width.
Each channel is allowed its own area and background cross section.\footnote{
The `area' parameter is usually chosen in the parameterization of the
resonance shape because it is proportional to the total number of
events in each channel.
It is less correlated with the width than the
product of branching fractions.}

\section{Beam energy measurements}
\label{sec:beam_energy}

The center-of-mass energy distribution $B_i(w)$
is critical for width and area
measurements.  We summarize here the concepts that are essential for
the following discussion.  More details can be found in
Refs.~\cite{armstrong:1993,mcginnis:2003}.

The revolution-frequency distribution of the antiprotons
is measured by detecting
the Schottky noise signal generated by the coasting beam.  The signal
is sensed by a 79-MHz longitudinal Schottky pickup and recorded on a
spectrum analyzer.  An accuracy of 0.05~Hz is achieved on a revolution
frequency of 0.63~MHz, over a wide dynamic range in intensity (60~dBm).

The beam is slightly bunched by an rf cavity operating at
$\fcav \sim \q{1.25}{MHz}$,
the second harmonic ($h=2$) of the revolution frequency.
The beam is bunched both for stability (ion clearing) and
for making the beam position monitors (BPMs) sensitive to a portion of
the beam.
Therefore, recorded orbits refer to particles bunched by the rf system,
and their revolution frequency is
\( \frf = \fcav / h \).
The bunched-beam revolution frequency~\frf\ 
is usually close to the average revolution
frequency of the beam.
Each orbit consists of 48
horizontal and 42 vertical readings.
As a result of hardware and
software improvements, these readings are much less noisy than
E760's~\cite{armstrong:1993}, as discussed later in the uncertainty
estimates.

From the BPM readings and the Accumulator lattice model,
we can accurately calculate differences \dL\
in the length of one orbit and another.
The main systematic uncertainties come from BPM calibrations, from bend-field drifts, and from neglecting second-order terms in the orbit length.
Using the dispersion function from the lattice model, the gains of the high-dispersion BPMs can be measured by varying the beam energy at constant magnetic field. They show calibration errors between 3\% and 15\%. Their systematic effect on~\dL\ is about 
0.03~mm. Contributions from calibration errors
in the low-dispersion BPMs are harder to evaluate, but they should be comparable.
Bend-field drifts (due to temperature variations, for instance) appear in the orbit-length calculation as changes in momentum.
For the \psip\ scans,
their contribution translates into an uncertainty in~\dL\ of about 0.04~mm.
The second-order terms depend on the derivatives of the vertical and horizontal orbit slope differences with respect to the reference orbit as well as the slopes of the reference orbit itself.
An explicit assessment of these terms is not possible, because there are not enough BPMs to measure slopes everywhere around the ring. Under reasonable assumptions, one would get an error of about 0.005~mm.
A test was performed to estimate the accuracy in~\dL.
The systematic uncertainty is evaluated by using a January 2000
\psip\ orbit to predict the length of a very different \psip\ orbit of
known length from August 1997, when the machine lattice was also quite
different.  The difference between the known length and the predicted
length is 0.05~mm out of 474~m.  Since orbits and lattices for the runs used in this analysis are much closer to each other, this is taken as the
systematic uncertainty in \dL\ from the beam-energy calculation for these runs.

The absolute length~$L$ of an orbit can be calculated from a reference orbit of length $L_0$:
\( L = L_0 + \dL \).
The calibration of~$L_0$ is
done by scanning a charmonium resonance
(the \psip\ itself in this analysis)
the mass of which is precisely known
from the resonant-depolarization method in \ee\
experiments~\cite{aulchenko:2003}.  For particles in the bunched
portion of the beam (rf bucket), the relativistic parameters~\betarf\
and~\gammarf\ are calculated from their velocity \( \vrf = \frf \cdot
L \), from which the center-of-mass energy~$w$ of the \pbarp\ system
is calculated:
\( \wrf = w(\frf,L) \equiv m \sqrt{2 \left( 1+\gammarf \right)} \).
(The superscript rf is omitted from orbit lengths
because they always refer to particles in the rf bucket.)  In the
charmonium region, this method yields good accuracies on~$w$.  For
instance, \( \partial w/ \partial f = \q{113}{keV/Hz} \)
(\q{38}{keV/Hz}) and \( \partial w/\partial L = \q{149}{keV/mm} \)
(\q{50}{keV/mm}) at the \psip\ (\jpsi).

For width and area determinations, energy differences are crucial, and
they must be determined precisely.  In our usual data-taking, where
we keep the beam near the central orbit of the Accumulator, a particular run is
chosen as the reference (subscript~0).  Energy differences between the
reference run and other runs in the scan (subscript~$i$), for
particles in the rf bucket, are simply
\begin{equation}
  \wrf_i - \wrf_0 =
  w(\frfi,L_0+\dL_i) - w(\frfz,L_0).
  \label{eq:dw_dL}
\end{equation}
Within the energy range of a resonance scan, these differences are
largely independent of the choice of~$L_0$.  For this reason, the
absolute energy calibration is irrelevant for width and area
measurements.  Only uncertainties coming from~\dL\ are considered.

Once the energy~$\wrf_i$ for particles in the rf bucket is known, the
complete energy distribution is obtained from the Schottky
spectrum using the relation between frequency differences and momentum
differences at constant magnetic field:
\begin{equation}
  \frac{\Delta p}{p} = - \frac{1}{\eta} \frac{\Delta f}{f},
  \label{eq:dpp}
\end{equation}
where~$\eta$ is the energy-dependent phase-slip factor of the machine,
which is one of the parameters governing synchrotron oscillations.
(The dependence of~$\eta$ on beam energy is chosen during lattice design, as described in Ref.~\cite{mcginnis:2003};
the variation of~$\eta$ within a scan can be neglected.)
In terms of the center-of-mass energy,
\begin{equation}
  w - \wrf_i = - \frac{1}{\eta}
    \frac{(\betarf_i)^2 (\gammarf_i) m^2}{\wrf_i}
    \frac{f - \frf_i}{\frf_i}.
  \label{eq:dw}
\end{equation}
Within a run,
rf frequencies,
beam-frequency spectra,
and BPM readings are updated every few minutes.
Frequency spectra are then translated into center-of-mass energy through
Eq.~\ref{eq:dw}, weighted by luminosity and summed, to obtain the
luminosity-weighted normalized energy spectra~$B_i(w)$ for each
data-taking run.

The phase-slip factor is usually determined from the synchrotron frequency.
In our case,
this determination has a 10\% uncertainty coming from the bolometric
rf voltage measurement~\cite{garzoglio:2004}.
At the \psip, the synchrotron-frequency method yields a phase-slip factor
\( \eta = 0.0216\pm 0.0022 \).

The resonance width and area are affected by a systematic error due to the
uncertainty in~$\eta$.  Usually, the resonance width and area are positively
correlated with the phase-slip factor.  A larger~$\eta$ implies a narrower energy spectrum, as described in Eq.~\ref{eq:dw}.  As a
consequence, the fitted resonance will more closely resemble the
measured excitation curve, yielding a larger resonance width.  For our
scan at the central orbit (stack~1), the 10\% uncertainty in~$\eta$
translates into a systematic uncertainty of about 18\% in the width
and 2\% in the area.

\section{Complementary scans}
\label{sec:compl_scans}

For precision measurements, one needs a better estimate of the
phase-slip factor or determinations that are independent of~$\eta$, or
both.
In E760, the `double scan'
technique was used~\cite{armstrong:1993}.
It yielded~$\eta$ with an uncertainty of 6\% at the \psip\
and width determinations largely independent of the phase-slip factor,
but it had the disadvantage of being operationally complex.

Here we describe a new
method of `complementary scans'
to achieve a similar precision on~$\eta$ and arbitrarily small correlations between resonance parameters and phase-slip factor;
the technique is also operationally simpler.
The resonance is scanned once on the central orbit, as described above.
A second scan is then performed at constant magnetic bend field
(most of stack~29, runs 5818--5831). 
The energy of the beam is changed by moving the
longitudinal stochastic-cooling pickups.  The beam moves away from
the central orbit, and the range of energies is limited but
appropriate for narrow resonances.

Since the magnetic field is
constant, beam-energy differences can be calculated independently
of~\dL, directly from the revolution-frequency spectra and the phase-slip
factor, according to Eq.~\ref{eq:dw}.
A pivot run is chosen (5827 in our case, subscript~$p$).
The rf frequency of this run is used as a reference to
calculate the energy for particles in the rf bucket in other runs.
These particles have revolution frequency~$\frf_i$ and the energy is calculated as follows:
\begin{equation}
 \wrf_i - \wrf_p = -\frac{1}{\eta}
 \frac{(\betarf_p)^2 (\gammarf_p) m^2}{\wrf_p}
 \frac{\frf_i - \frf_p}{\frf_p}.
 \label{eq:dw_eta}
\end{equation}
For the scan at constant magnetic field,
this relation is used instead of Eq.~\ref{eq:dw_dL}.
Once
the energy for particles at $\frf_i$ is known, the full energy spectrum within each run is obtained from Eq.~\ref{eq:dw}, as usual.\footnote{
For the constant-field scan,
the energy distributions may be obtained directly from the pivot energy
by calculating \( w - \wrf_p \),
instead of using Eq.~\ref{eq:dw_eta} first and then Eq.~\ref{eq:dw}.
The two-step procedure is chosen because it is faster to rescale the energy spectra than to re-calculate them from the frequency spectra when fitting for~$\eta$.
Numerically,
the difference between the two calculations is negligible
(less than 0.2~keV).
Moreover, the two-step procedure exposes how the width depends on~$\eta$.}

Using this alternative energy measurement, the width and area determined
from scans at constant magnetic field are negatively correlated
with~$\eta$.  The increasing width with increasing~$\eta$ is still
present, as it is in scans at nearly constant orbit.  But the dominant
effect is that a larger~$\eta$ brings the energy points in the
excitation curve closer to the pivot point, making the width smaller.  In the
case of stack~29, a 10\% increase in~$\eta$ implies a -10\% variation
in both width and area.

\begin{figure}[h]
  \centering
    \resizebox{4in}{!}{\includegraphics{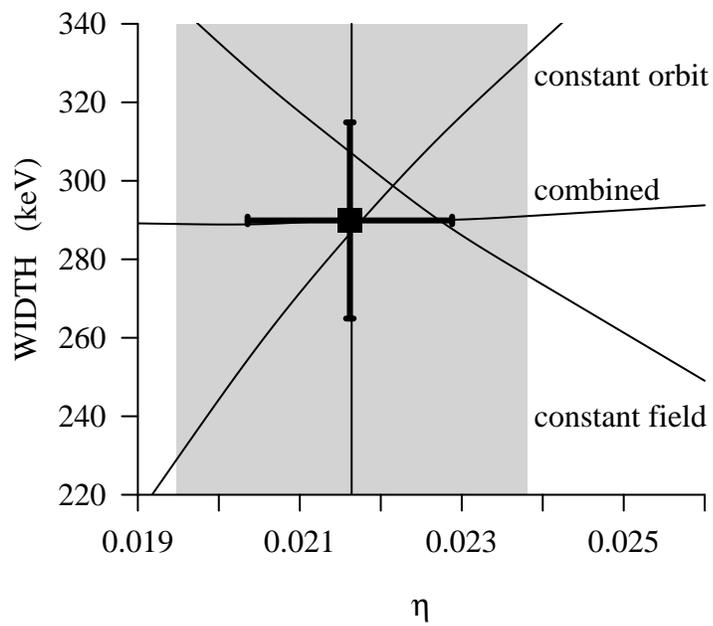}}
    \caption{$\Gamma$ dependence on~$\eta$ for stacks~1 and~29,
      and from their combination, when the phase-slip factor is a fixed parameter.
      The result of the global fit with free~$\eta$ is represented by the cross.
      The value of the phase-slip factor from the
      synchrotron-frequency measurement (vertical line) and its uncertainty (gray band) are also shown.}
    \label{fig:crossing}
\end{figure}

The different dependence of the width on~$\eta$ in the
two separate scans is shown as two crossing curves in Figure~\ref{fig:crossing}.
(Statistical errors, $\pm \q{36}{keV}$ for both curves, are not shown.)
The constant-orbit and the constant-field scan can be combined.
The resulting width has a dependence on~$\eta$ that is intermediate between the two.
An appropriate luminosity distribution can make the resulting curve practically horizontal.
The combined measurement is dominated by the statistical uncertainty
($\pm \q{25}{keV}$, in this case; not shown in the plot).

Moreover,
thanks to this complementary behavior,
the width, area and phase-slip factor
can be determined in a maximum-likelihood fit where~$\eta$ is also a free parameter.
Errors and correlations are then obtained directly from the fit.

\section{Results}
\label{sec:results}

\begin{figure}[h]
  \centering
    \resizebox{5in}{!}{\includegraphics{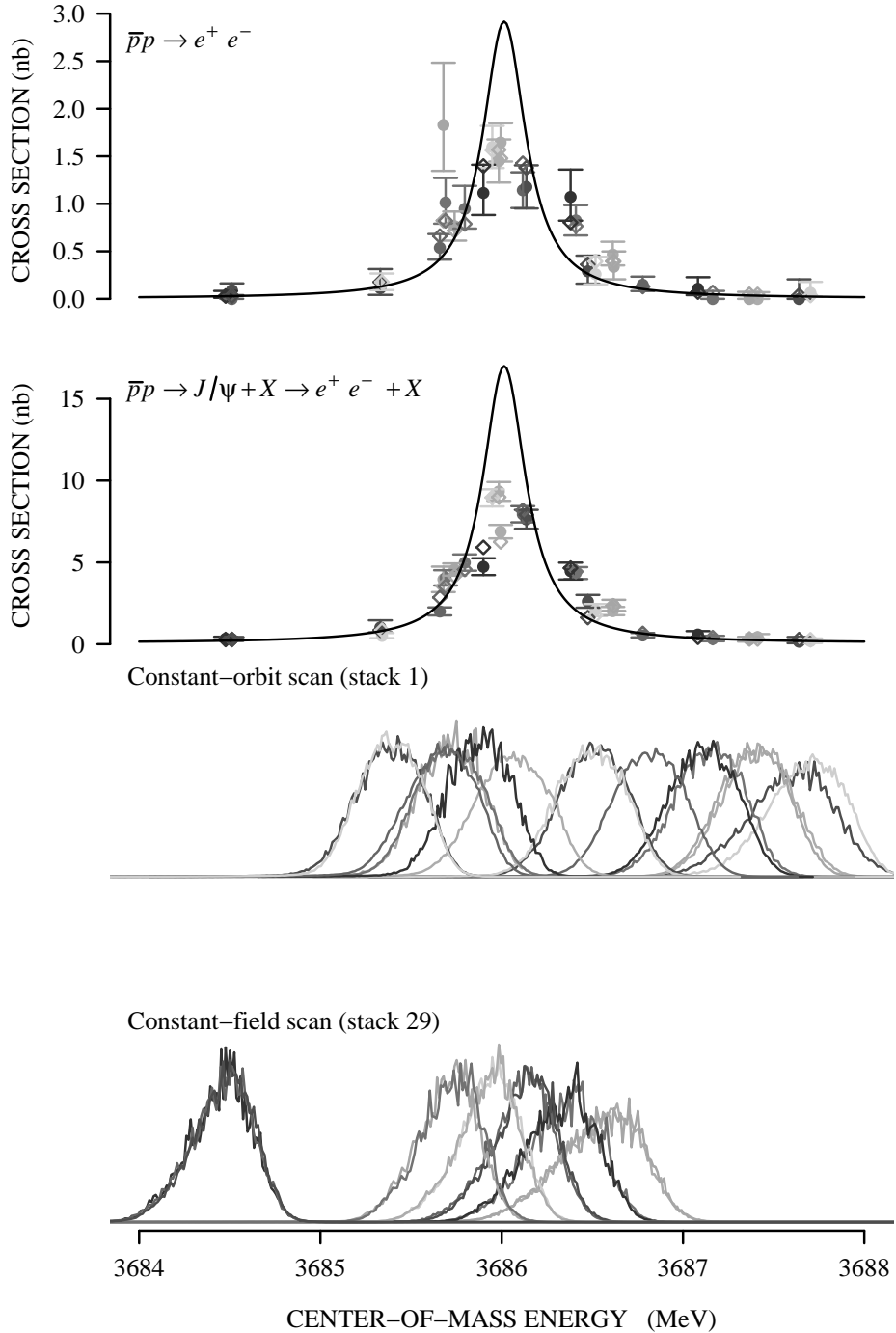}}
  \caption{\psip\ resonance scans:
    the observed cross section for each channel (filled dots);
    the expected cross section from the fit (open diamonds);
    the `bare' resonance curves~\sBW\ from the fit (solid lines).
    The two bottom plots show the normalized energy distributions~$B_i$.}
  \label{fig:both}
\end{figure}

Both channels in both scans are fitted simultaneously,
leaving the phase-slip factor as a free parameter.
The energy distributions are rescaled
according to Eq.~\ref{eq:dw} for the `constant-orbit' scan
and Eqs.~\ref{eq:dw} and~\ref{eq:dw_eta} for the `constant-field' scan.
The log-likelihood function is
\( \log(\lik) =
   \sum_i \left[ \log P(\muee_i,\Nee_i) + \log P(\muX_i,\NX_i) \right] \).
For each channel, the mean numbers of events $\muee_i$ and $\muX_i$ are evaluated according to Eq.~\ref{eq:mu}.
We monitor the efficiency of each data run and the efficiencies vary less than 0.4\% over a single scan.
Both scans have the same \ee\ efficiency
\( \effee = 0.413 \pm 0.015 \).
For the \psinc\ channel,
the constant-field scan efficiency is
\( \effXcf = 0.402 \pm 0.011 \);
differences in
detection efficiency between the two scans
are accounted for by the parameter \((\reff)\).
They are due to a different configuration of the tracking system,
which does not affect the \ee\ channel.
The likelihood maximization was performed within
the~\code{R} package~\cite{R:2005} and
crosschecked with the \code{Minuit} code~\cite{james:1994}.
The results of the fit are shown in 
Figure~\ref{fig:crossing}, Figure~\ref{fig:both}, and Table~\ref{tab:results}.\footnote{The \psip\ mass from Ref.~\cite{aulchenko:2003} is used for the absolute calibration of~$L_0$. The value of~$M$ in Table~\ref{tab:results} is not an independent measur
ement.}

The fitted value of~$\eta$ in Table~\ref{tab:results}
is consistent with the one determined from
the synchrotron frequency (Section~\ref{sec:beam_energy}).
The relative uncertainty in the phase-slip
factor (6\%) is equal to that from the E760 double scans~\cite{armstrong:1993}.

Possible statistical and systematic sources of uncertainty in the
width and area are considered.
As discussed in Section~\ref{sec:beam_energy},
each beam spectrum is a luminosity-weighted sum of individual energy
distributions within each run.
Statistical fluctuations of the BPM readings
produce random variations of the
measured~\dL\ which systematically widen the beam spectrum,
making the resonance width narrower.
The BPM noise is evaluated from portions of runs with no energy drifts and
its standard deviation is 0.02~mm for both stacks.
As a result of hardware and software improvements,
this is much lower than the E760 value, 0.2~mm~\cite{armstrong:1993}.
The effect on width and area due to BPM noise
is larger for small beam widths and for
runs with no energy drifts.
In the worst case,
it translates into a systematic uncertainty of $<8$~keV in the width and
$<2$~meV in the area.
We do not correct for this systematic, but uncertainties
are assigned to the results of 4~keV and 1~meV, respectively.

The systematic uncertainties in the luminosity (2.5\%) and \ee\
efficiency (3.6\%) directly affect the area, but not the width.
They are added to obtain an uncertainty of 6.1\% or 35~meV.

The absolute energy calibration does not influence the resonance
width and area.
Instead, a systematic error in the~\dL\ determination has the
following effects:
it shifts all runs in stack~1;
it shifts stack~29 with respect to stack~1.
The systematic uncertainty of 0.05~mm discussed in
Section~\ref{sec:beam_energy} translates into \q{<1}{keV} for the width
and \q{<1}{meV} for the area,
and it is therefore neglected.
No systematic uncertainties are therefore introduced by
combining the two scans.
The pivot run in the constant-field scan
(Section~\ref{sec:compl_scans}) is taken near the central
orbit and has a small~\dL,
so that its energy relative to the constant-orbit scan can be
calculated accurately from Eq.~\ref{eq:dw_dL}.

Our final results are the following:
\begin{eqnarray}
  \Gamma & = & \resG \\
  & & \nonumber \\
  \GGG & = & \resA.
\end{eqnarray}

\section{Discussion}

\begin{figure}[h]
  \centering
    \resizebox{4in}{!}{\includegraphics{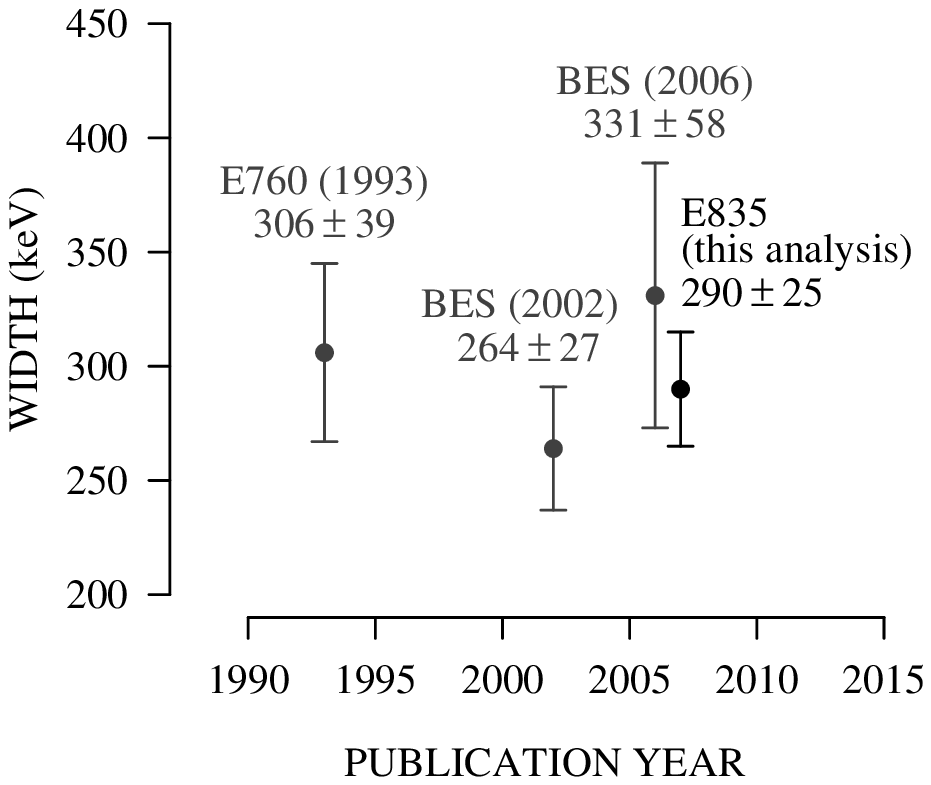}}
    \caption{Recent measurements of the \psip\ width.}
    \label{fig:compare}
\end{figure}

A comparison between this width measurement and those of
E760~\cite{armstrong:1993} and BES~\cite{bai:2002,ablikim:2006} is shown in
Figure~\ref{fig:compare}.  All three values are consistent.  The
E835 measurement is the most precise.
Our measurement of $(\GGG)$ is also compatible,
but much more precise,
than that reported by BABAR,
\( \GGG = \q{0.70 \pm 0.17 \pm 0.03}{eV} \)~\cite{aubert:2006}.

This new method of complementary scans
can be applied to future experiments
for the direct determination of narrow resonance widths
in antiproton-proton annihilations
(such as PANDA at the future FAIR facility in Darmstadt).
If one performs a scan at constant orbit and a scan at constant magnetic
field in conditions similar to those in the Antiproton Accumulator,
the uncertainty is mainly statistical.
Moreover, by appropriately
choosing the relative luminosities and energies of the two scans, one can make the
width almost uncorrelated with the phase-slip factor, as in the E835 case
discussed in this paper.

\section{Acknowledgements}

We gratefully acknowledge the support of the Fermilab staff and
technicians and especially the Antiproton Source Department of the
Accelerator Division and the On-line Department of the Computing
Division.
We also wish to thank the INFN and University technicians and
engineers from Ferrara, Genoa, Turin and Northwestern for their
valuable work.
This research was supported by the Italian
\emph{Istituto Nazionale di Fisica Nucleare} (INFN)
and the US Department of Energy.



\pagebreak

\pagebreak

\begin{table}[h]
  \centering
    \begin{tabular}{rrrrr}
      \multicolumn{1}{c}{Run} &
      \multicolumn{1}{c}{Energy} &
      \multicolumn{1}{c}{Luminosity} &
      \multicolumn{1}{c}{\ee\ events} &
      \multicolumn{1}{c}{\(\jpsi+X\) events} \\
      \multicolumn{1}{c}{$i$} & $\wrf_i$ (MeV)  & $\Lum_i$ (nb$^{-1}$)  & $\Nee_i$ & $\NX_i$ \\
      \hline
      5006 &  3687.585 &   15.20 &   0    &  1   \\
      5009 &  3687.632 &   37.20 &   1    &  3   \\
      5012 &  3687.373 &   44.40 &   0    &  8   \\
      5013 &  3687.343 &   42.20 &   0    &  5   \\
      5015 &  3687.121 &   37.20 &   0    &  5   \\
      5016 &  3687.080 &   45.10 &   2    &  11  \\
      5019 &  3686.760 &   80.80 &   5    &  18  \\
      5022 &  3686.471 &   41.58 &   5    &  45  \\
      5023 &  3686.453 &   37.15 &   4    &  32  \\
      5025 &  3686.012 &   98.56 &   67   &  280 \\
      5027 &  3685.678 &   15.88 &   12   &  26  \\
      5028 &  3685.667 &   45.37 &   19   &  76  \\
      5029 &  3685.848 &   43.44 &   20   &  85  \\
      5031 &  3685.643 &   80.99 &   18   &  67  \\
      5036 &  3685.338 &   21.22 &   1    &  9   \\
      5038 &  3685.334 &   61.43 &   4    &  13  \\
      \hline
      &           &  747.72 &   158  &  684 \\
    \end{tabular}
  \caption{Stack~1 data.}
  \label{tab:stack01}
\end{table}

\begin{table}[h]
  \centering
    \begin{tabular}{rrrrr}
      \multicolumn{1}{c}{Run} &
      \multicolumn{1}{c}{Energy} &
      \multicolumn{1}{c}{Luminosity} &
      \multicolumn{1}{c}{\ee\ events} &
      \multicolumn{1}{c}{\(\jpsi+X\) events} \\
      \multicolumn{1}{c}{$i$} & $\wrf_i$ (MeV)  &
       $\Lum_i$ (nb$^{-1})$   & $\Nee_i$ & $\NX_i$\\
      \hline
      5818 & 3686.674 &  77.81 & 15  & 65       \\
      5819 & 3686.701 &  50.01 & 7   & 49       \\ 
      5821 & 3686.422 &  79.13 & 27  & 142      \\
      5822 & 3686.427 &  40.63 & 18  & 75       \\
      5824 & 3686.126 &  78.34 & 37  & 257      \\
      5825 & 3686.138 &  55.50 & 27  & 175      \\
      5827 & 3685.922 &  78.80 & 52  & 291      \\
      5828 & 3685.922 &  68.44 & 41  & 264      \\
      5830 & 3685.633 &  79.07 & 25  & 149      \\
      5831 & 3685.643 &  48.52 & 19  & 100      \\
      5833 & 3684.455 &  79.31 & 1   & 11       \\ 
      5834 & 3684.450 &  78.35 & 0   & 10       \\
      5837 & 3684.451 &  78.68 & 3   & 10       \\ 
      \hline
      &          & 892.59 & 272 & 1598     \\
    \end{tabular}
  \caption{Stack~29 data.}
  \label{tab:stack29}
\end{table}

\begin{table}[h]
  \centering
\begin{tabular}{rlrrrrrrrr}
\multicolumn{2}{c}{Parameter} &
\multicolumn{1}{c}{Value} &
\multicolumn{7}{c}{Correlation matrix} \\
   &           &                   &   2  &   3  &    4  &    5  &    6  &     7 &     8 \\
\hline
 1 & $M$ (MeV)       & $3686.111 \pm 0.009$ & 0.02 & 0.35 & -0.07 & -0.64 & -0.23 &  0.10 & -0.17 \\
 2 & $\Gamma$ (keV) & $290 \pm 25$         &      & 0.35 & -0.29 &  0.02 & -0.07 &  0.05 & -0.52 \\
 3 & \GGG\ (meV)     & $579 \pm 38$         &      &      & -0.44 & -0.48 & -0.28 & -0.58 & -0.27 \\
 4 & \sbkg(\ee) (pb)      & $3 \pm 6$            &      &      &       &  0.09 &  0.07 &  0.28 &  0.17 \\
 5 & $\eta \ (10^{-4})$ & $216\pm 13$ &      &      &       &       &  0.54 & -0.20 &  0.17 \\
 6 & \reff\ (\%) & $73 \pm 4$           &      &      &       &       &       & -0.29 &  0.02 \\
 7 & $\frac{\mathrm{Area}(\psinc)}{\mathrm{Area}(\ee)}$ & $5.81\pm 0.35$   &      &      &       &       &       &       & -0.17 \\
 8 & \sbkg(\psinc) (pb)   & $65\pm 19$           &      &      &       &       &       &       &       \\
\multicolumn{2}{l}{\(\left[\log(\lik)\right]_\mathrm{max}\)} & -170.95 & \multicolumn{7}{c}{} \\
\multicolumn{2}{l}{Goodness-of-fit tests:} & \multicolumn{8}{c}{} \\
   & log-likelihood ratio & 68.8 / 50 ($P = 4.0\%$) & \multicolumn{7}{c}{} \\
   & $\chi^2$ / d.o.f.    & 68.3 / 50 ($P = 4.3\%$) & \multicolumn{7}{c}{} \\
  \end{tabular}
  \caption{Summary of the results of the maximum-likelihood fit.}
  \label{tab:results}
\end{table}

\pagebreak

\begin{thebibliography}{00}




\bibitem{bai:2002} BES Collaboration, J.~Z.~Bai et al., Phys.\ Lett.\
  B \textbf{550} (2002)~24.

\bibitem{ablikim:2006} BES Collaboration, M.~Ablikim et al.,
  Phys.\ Rev.\ Lett.\ \textbf{97} (2006)~121801.

\bibitem{aubert:2006} BABAR Collaboration, B.~Aubert et al., Phys.\
  Rev.\ D \textbf{73} (2006)~012005.

\bibitem{armstrong:1993} E760 Collaboration, T.~A.~Armstrong et al.,
  Phys.\ Rev.\ D \textbf{47} (1993)~772.
  We note that an error in the E760 calculation of initial-state radiation
  resulted in an overestimate of the \psip\ and \jpsi\ widths. The corrected
  E760 results are
  \( \Gamma_\mathrm{E760} = \q{287 \pm 37 \pm 16}{keV} \) for the \psip\ and
  \( \Gamma_\mathrm{E760} = \q{91 \pm 11 \pm 6}{keV} \) for the \jpsi\
  (C.~M.~Ginsburg and P.~A.~Rapidis, private communication).

\bibitem{garzoglio:2004} E835 Collaboration, G.~Garzoglio et al.,
  Nucl.\ Instrum.\ Methods A \textbf{519} (2004)~558.

\bibitem{mcginnis:2003} D.~P.~McGinnis, G.~Stancari, and
  S.~J.~Werkema, Nucl.\ Instrum.\ Methods A \textbf{506} (2003)~205.

\bibitem{kennedy:1992} D.~C.~Kennedy, Phys.\ Rev.\ D \textbf{46}
  (1992)~461.

\bibitem{andreotti:2005} E835 Collaboration, M.~Andreotti et al.,
  Phys.\ Rev.\ D \textbf{71} (2005)~032006.




\bibitem{aulchenko:2003} V.~M.~Aulchenko et al., Phys.\ Lett.\ B
  \textbf{573} (2003)~63.

\bibitem{R:2005} R Development Core Team (2005).
   R: A language and environment for statistical computing.
   R Foundation for Statistical Computing, Vienna, Austria.
   ISBN 3-900051-07-0, \url{<http://www.R-project.org>}.

\bibitem{james:1994} F.~James,
   \code{Minuit}: Function Minimization and Error Analysis,
   Version 94.1,
   CERN Program Library Long Writeup~D506.
   
\end{thebibliography}
\end{document}